\documentclass{edm_article}

\usepackage{url} 
\usepackage{booktabs}
\usepackage{caption}
\usepackage{multirow}
\usepackage{subfig}

\newtheorem{assumption}{Assumption}

\begin{document}

\title{LANA: Towards Personalized Deep Knowledge Tracing Through Distinguishable Interactive Sequences
}
%
\numberofauthors{6}
\author{
\alignauthor
Yuhao Zhou\\
    \affaddr{Sichuan University}\\
    \affaddr{Chengdu, China}\\
    \email{sooptq@gmail.com}\\
\alignauthor
Xihua Li\\
    \affaddr{Tencent Inc.}\\
    \affaddr{Chengdu, China}\\
    \email{lixihua9@126.com}\\
\alignauthor
Yunbo Cao\\
    \affaddr{Tencent Inc.}\\
    \affaddr{Beijing, China}\\
    \email{yunbocao@tencent.com}\\
\and
\alignauthor
Xuemin Zhao\\
    \affaddr{Tencent Inc.}\\
    \affaddr{Chengdu, China}\\
    \email{xueminzhao@tencent.com}\\
\alignauthor
Qing Ye\\
    \affaddr{Sichuan University}\\
    \affaddr{Chengdu, China}\\
    \email{fuyeking@gmail.com}\\
\alignauthor
Jiancheng Lv\\
    \affaddr{Sichuan University}\\
    \affaddr{Chengdu, China}\\
    \email{lvjiancheng@scu.edu.cn}
}

\maketitle


\begin{abstract}
In educational applications, \textit{Knowledge Tracing} (KT), the problem of accurately predicting students' responses to future questions by summarizing their knowledge states, has been widely studied for decades as it is considered a fundamental task towards adaptive online learning. Among all the proposed KT methods, Deep Knowledge Tracing (DKT) and its variants are by far the most effective ones due to the high flexibility of the neural network. However, DKT often ignores the inherent differences between students (e.g. memory skills, reasoning skills, ...), averaging the performances of all students, leading to the lack of personalization, and therefore was considered insufficient for adaptive learning. To alleviate this problem, in this paper, we proposed \textbf{L}eveled \textbf{A}ttentive K\textbf{N}owledge Tr\textbf{A}cing (LANA), which firstly uses a novel student-related features extractor (SRFE) to distill students' unique inherent properties from their respective interactive sequences. Secondly, the pivot module was utilized to dynamically reconstruct the decoder of the neural network on attention of the extracted features, successfully distinguishing the performance between students over time. Moreover, inspired by Item Response Theory (IRT), the interpretable Rasch model was used to cluster students by their ability levels, and thereby utilizing leveled learning to assign different encoders to different groups of students. With pivot module reconstructed the decoder for individual students and leveled learning specialized encoders for groups, personalized DKT was achieved. Extensive experiments conducted on two real-world large-scale datasets demonstrated that our proposed LANA improves the AUC score by at least 1.00\% (i.e. EdNet~$\uparrow 1.46\%$ and RAIEd2020~$\uparrow 1.00\%$), substantially surpassing the other State-Of-The-Art KT methods.
\end{abstract}
%
\section{Introduction}
\begin{figure}[t]
    \centering
    \includegraphics[width=0.9\linewidth]{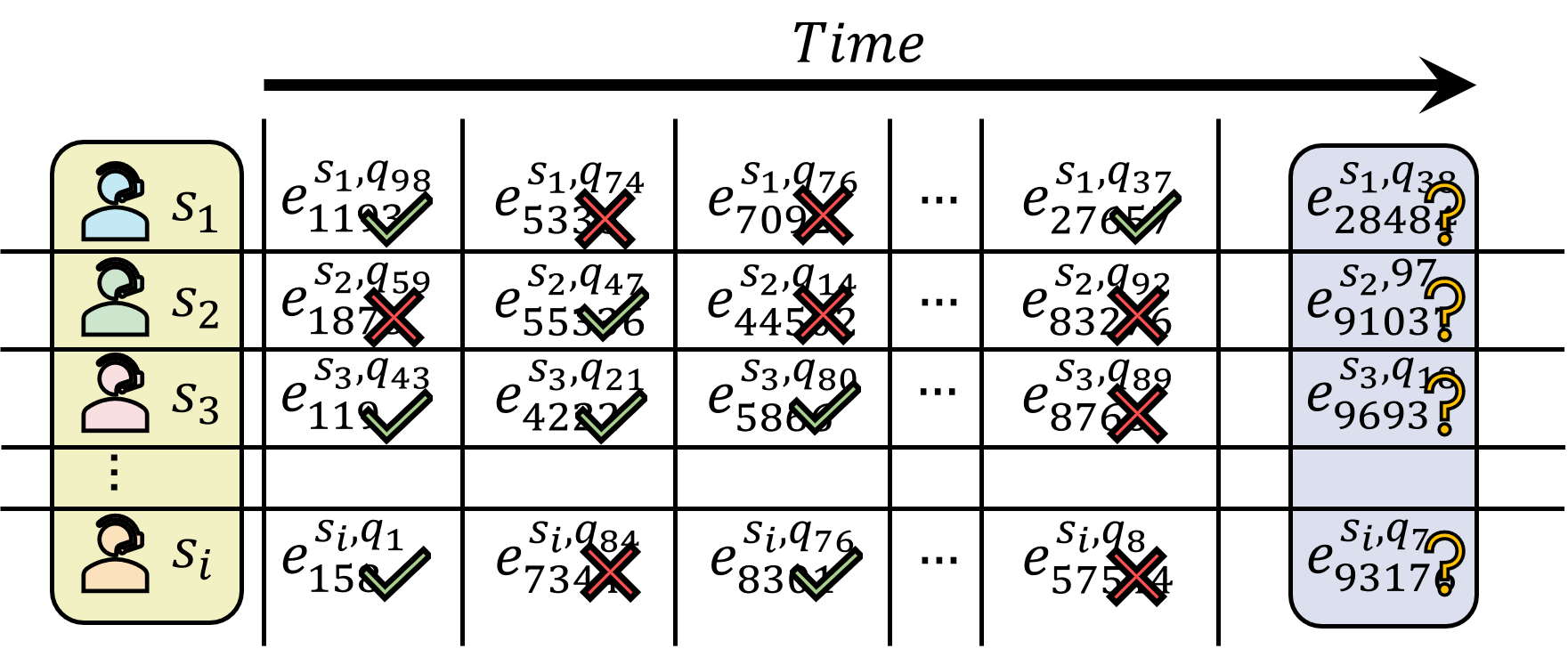}
    \caption{A sample \textit{Knowledge Tracing} (KT) process. Given each student's exercising activities in the past, KT aims to accurately predict the probability of whether each student is capable of correctly answering the next question.}
    \label{fig:kt-example}
\end{figure}
\textit{Knowledge Tracing} (KT) aims to accurately retrieve students' knowledge states at a certain time by his past sequential exercising interactions. To evaluate KT's performance, it is asked to predict the correctness of students' future exercises with the retrieved knowledge states as Figure~\ref{fig:kt-example}. Traditionally, KT was regarded as a sequential behavior mining task~\cite{jalal2019students, shang2017searching}, and therefore various methods established models with the theory of bayesian probability (BKT~\cite{corbett1994knowledge}) and psycho-statistics (IRT~\cite{gonzalez2014general}), providing excellent interpretability and good performance. Nevertheless, recently proposed Deep Knowledge Tracing (DKT)~\cite{piech2015deep} and its variants~\cite{pandey2019self, ghosh2020context, pandey2020rkt, choi2020towards, shin2020saint+} significantly outperform other KT methods in metrics using Recurrent Neural Network (RNN) and Long Short Term Memory     (LSTM~\cite{hochreiter1997long}). However, DKT distinctly lacks personalization for students compared to BKT and IRT, which are capable of separately training unique models for each student, while DKT only trains a unified model for all students due to massive training data and abundant computing resources required by deep learning. Hence, DKT weakly reflects the large inherent property (i.e. memory skills, reasoning skills, or even guessing skills) gaps between students.

\textbf{Is it possible to bring personalization back to DKT?} To answer this question, a natural and straightforward thought is to treat KT task as an exercise-recommending task in which students are the target users. However, a typical recommendation system usually requires much targets' side information (e.g. age, sex, education, ...) to help the model realize the differences between users, while KT could only provide little. Moreover, the side information in the recommendation system is often fixed or changing gradually, instead of being dynamic in KT (e.g. memory skills could be reinforced over time). As a result, employing an excellent recommendation system in KT will mostly fail to properly cluster different students due to insufficient student-related features, resulting in an unsatisfactory personalizing experience.

Motivated by the observation that the proactive behavior sequence (i.e. interactive sequences) of each individual is unique and changeable over time, we argue that the minimal personalized unit in KT is ``a student at a certain time $t_i$'' instead of just ``a student'', and \textbf{student's inherent properties at time $t_i$ can be represented by his interactive sequences around time $t_i$}. In such a way, these student-related features could tremendously help personalize the KT process since they could be used to identify different students at different stages. Consequently, in our proposed \textbf{L}eveled \textbf{A}ttentive K\textbf{N}owledge Tr\textbf{A}cing (LANA), unique student-related features are distilled from students' interactive sequence by a Student-Related Features Extractor (SRFE). Moreover, inspired by BKT and IRT that assign completely different models to different students, LANA, as a DKT model, successfully achieves the same goal in a different manner. Detailedly, instead of separately training each student a model like BKT and IRT, \textbf{LANA learns to learn correlations between inputs and outputs on attention of the extracted student-related features}, and thus becomes transformable for different students at different stages. More specifically, the transformation was accomplished using pivot module and leveled learning, where the former one is a model component that seriously relies on the SRFE, and the latter one is a training mechanism that specializes encoders for groups with interpretable Rasch model defined ability levels. Formally, the LANA can be represented by:
\begin{equation}\label{eq:kt-def3}
    \overbrace{r_t^{s_i} \sim (f(p_t^{s_i}))(h_t^{s_i})}^{Adaptive~by~Pivot~Module},~~\underbrace{p_t^{s_i} \sim k(h_t^{s_i}),~~h_t^{s_i} \sim g(h_{<t}^{s_i}, I_{<t}^{s_i})}_{Adaptive~by~Leveled~Learning},
\end{equation}
where $h_t^{s_i}$ is referred as student $s_i$'s knowledge state at time $t$, $I_{<t}^{s_i} \sim \{e_{<t}^{s_i}, r_{<t}^{s_i}\}$ is referred as the sequential exercising interactions of student $s_i$ before time $t$, and finally $f(\cdot)$ (decoder), $g(\cdot)$ (encoder) and $k(\cdot)$ (SRFE) are three main modules that LANA seeks to learn.

The main contributions of this paper were summarized as:

\begin{enumerate}
    \item To the best of our knowledge, LANA was the first proposal to distill student-related features from their respective interactive sequences by a novel Student-Related Features Extractor (SRFE), exceedingly reducing the difficulty of achieving personalized KT.
    \item With distilled unique student-related features, novel pivot module and leveled learning were utilized in LANA to make the whole model transformable for different students at different stages, bringing strong adaptability to the DKT domain.
    \item Extensive experiments were conducted on two real-world large-scale KT datasets in comparison with other State-Of-The-Art KT methods. The results demonstrated that LANA outperforms any other KT methods substantially. Ablation studies were also performed to investigate the impact of different key components in LANA. The source code and hyper-parameters of experiments are open-sourced for reproducibility\footnote{\url{https://github.com/Soptq/LANA-pytorch}}.
    \item Visualizations of the intermediate features suggest extra impacts of LANA, such as learning stages transfer and learning path recommendation.
\end{enumerate}

The remainder of this paper was structured as follows. The background of KT was introduced in Section~\ref{section:two}. The problem setup of KT was formally stated in Section~\ref{section:three}. The comprehensive implementation of the proposed LANA was presented in Section~\ref{section:four}. Experiments and results analysis was demonstrated in Section~\ref{section:five}. Finally, Section~\ref{section:six} concluded this paper.

\section{Literature Review}
\label{section:two}

\subsection{Background}

With the rapid development of internet-related facilities, many online education systems~\cite{woolf2010building, anderson2014engaging}, such as Massive Online Open Courses (MOOC), Tencent Classroom~\footnote{\url{https://ke.qq.com/}} and Fudao QQ~\footnote{\url{https://fudao.qq.com/}}, are gaining more and more attention for allowing students to be educated by a wide range of high-quality courses across the world at any time and anywhere, potentially saving students extensive time and thereby improving their learning efficiency extraordinarily~\cite{oakley2019we}. Moreover, influenced by the pandemic in 2020, the number of students learning online has already surpassed 180 million. In addition with over 16 thousand courses and 67 degrees provided by over 950 universities, it is fair to say online education is making knowledge truly accessible for everyone in the world.

However, despite the prevalence and convenience of online education, it originally suffers from the inability to accurately track students' knowledge states, and thereby failing to provide personalized learning suggestions~\cite{kuh2011piecing} and perceive whether students have reached their objectives~\cite{grossman2011transfer}. In other words, in comparison with offline education at school, online education offers very little opportunity for students to proactively exporting their thoughts. Accordingly, the knowledge states of students learning online can be easily overlooked, resulting in a potential risk to students' future developments. Being motivated by the fact that students' current knowledge states are highly relevant to their previous learning activities, \textit{Knowledge Tracing} (KT)~\cite{corbett1994knowledge} was therefore introduced to address these problems by modeling students' knowledge states with their past sequential exercising interactions, and is commonly evaluated by predicting students' responses to future exercises:
\begin{equation}\label{eq:kt-def1}
    r_t^{s_i} \sim f(h_t^{s_i}),~~h_t^{s_i} \sim g(I_{<t}^{s_i}),
\end{equation}
where the meaning of each symbol is consistent with Equation~\ref{eq:kt-def3}. In addition, a sample KT process was illustrated in Figure~\ref{fig:kt-example}, where every student has sequential exercising interactions, each represents student $s_i$ correctly/incorrectly solved question $q_j$ at time $t$. Finally, a KT method is asked to accurately predict the probability of these students correctly solving the next question in the future.

\subsection{Related Work}

Extensive approaches~\cite{romero2010educational, desmarais2012review} were proposed to better model students' knowledge states from earlier Bayesian Knowledge Tracing (BKT) to recent Deep Knowledge Tracing (DKT). In the literature, BKT~\cite{corbett1994knowledge, d2008more} mainly assumes each student's knowledge state $h_t^{s_i}$ as a binary array with each bit representing whether student $s_i$ mastered a specific concept. Detailedly, BKT utilizes each student's sequential interactions to extract each question's latent concepts, and then solves $f(\cdot)$ and $g(\cdot)$ in Equation~\ref{eq:kt-def1} with a Hidden Markov Model~\cite{rabiner1986introduction} and 4 probabilistic parameters: initial mastery probability of a concept, transitioning probability from non-mastery to mastery of a concept, probability of guessing correct answers of questions related to a concept and the slipping probability. However, although BKT owns strong interpretability, the original work could only handle the scenario where each question only has one single concept, which is supposed to be not very practical in real-life situations. Hence, much later work was proposed to relax this assumption~\cite{xu2011using}. Meanwhile, inspired by psycho-statistics, Item Response Theory (IRT)~\cite{van2013handbook, wilson2016back, gonzalez2014general} was proposed to trace students' knowledge states with only 2 parameters (i.e. Rasch model): student ability and question difficulty, and thereby drastically simplified the equation-solving process. Moreover, each question in IRT was considered to have multiple concepts, assuring the effectiveness of IRT in real-world applications. Nevertheless, IRT freezes the parameters of the model once it was well trained and fitted to the data. Videlicet, performing online knowledge states updating is considerably complex for IRT. On the other hand, while $f(\cdot)$ and $g(\cdot)$ in BKT and IRT were both manually designed by specialists, recently DKT~\cite{piech2015deep} was proposed to utilize RNN/LSTM~\cite{hochreiter1997long} to automatically learn their representations. The idea of DKT is that the knowledge state of a student at the current time, namely $h_t^{s_i}$, can be inferred from his knowledge state history $h_{<t}^{s_i}$ and previous sequential exercising interactions $I_{<t}^{s_i}$:
\begin{equation}\label{eq:kt-def2}
    r_t^{s_i} \sim f(h_t^{s_i}),~~h_t^{s_i} \sim g(h_{<t}^{s_i}, I_{<t}^{s_i}), 
\end{equation}
where $g(\cdot)$ and $f(\cdot)$ are also known as the \textit{encoder} and the \textit{decoder} in DKT's context. Particularly, SAKT~\cite{pandey2019self}, to our knowledge, was the first work that replaces RNN/LSTM module with attention mechanism~\cite{vaswani2017attention}, substantially surpassing other KT methods in both speed and metrics. Moreover, SAINT+~\cite{shin2020saint+} stepped further on the basis of SAKT, by leveraging the whole transformer to learn the hidden patterns of students' interactive sequences, successfully achieving SOTA on EdNet~\cite{choi2020ednet} dataset, which significantly scaled the model up. AKT~\cite{ghosh2020context}, on the other hand, sought the possibility of improving the existing SAKT model. Specifically, AKT introduced a Monotonic Attention Mechanism (which would be referred to as the Vanilla Memory Attention (VMA) module in the latter of this paper) to take students' ``forgetting'' behaviors into account. Although these KT methods were all capable of efficiently tracking students' knowledge states over time, they were hard to provide personalized analysis and suggestions for each individual due to the unified model parameters for every student. In fact, from the perspective of these KT methods, all students are the same and share the same inherent properties (e.g. memory skills, reasoning skills or even guessing skills, ...). Furthermore, the lack of adaptability in DKT is actually a setback~\cite{pardos2010modeling, yudelson2013individualized} since both BKT and IRT have such an ability by assigning different models to different students. Consequently, LANA was therefore proposed, to bringing personalization back to DKT.

\begin{figure*}[!t]
    \centering
    \includegraphics[width=0.85\linewidth]{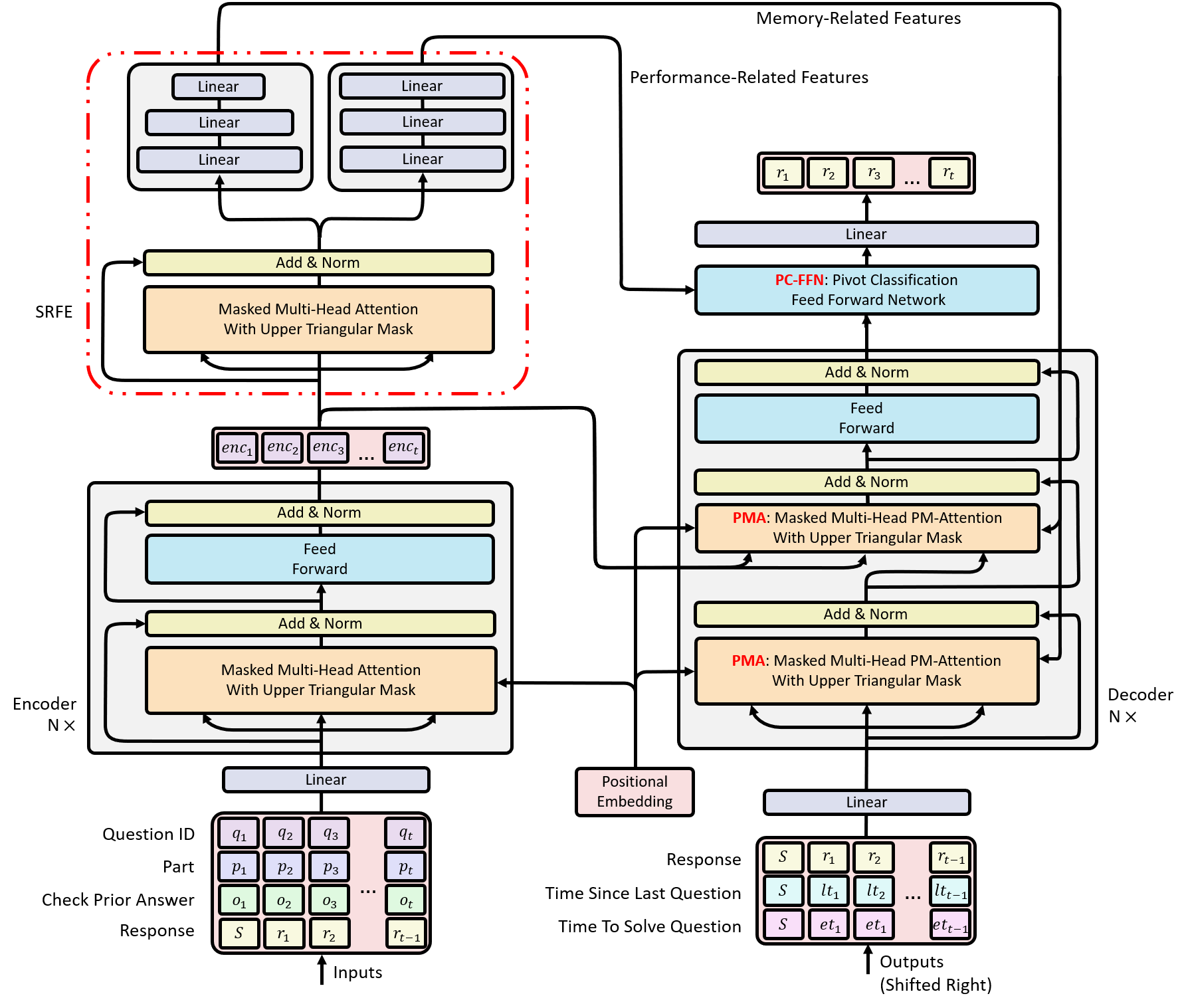}
    \caption{The overall model architecture of LANA. There are mainly three differences compared to vanilla transformer-based KT method~\cite{choi2020towards, shin2020saint+}: I. Modifications to the basic transformer model. II. Introduced SRFE and III. Introduced Pivot Memory Attention (PMA) Module and Pivot Classification Feed Forward Network (PC-FFN) Module, which collectively referred to as pivot module.}
    \label{fig:lana-arch}
\end{figure*}

\section{Problem Setup}
\label{section:three}

\subsection{Knowledge Tracing Problem}
Given a sequence of interactions $S_{t_0, t_1}^{s_i} = \{I_{t}^{s_i}|t_0 < t < t_1\}$, where $I_t^{s_i} = (e_{t}^{s_i,q_j}, r_t^{s_i})$, $e_{t}^{s_i,q_j}$ represents the activity of student $s_i \in \mathbb{N}^{+}$ answering question $q_j \in \mathbb{N}^{+}$ at discrete time step $t \in \mathbb{N}^{+}$, and $r_t^{s_i} \in \{0, 1\}$ represents the correctness of student $s_i$'s answer to $q_j$ at time $t$, the KT problem aims to predict the probability of $r_{t+1}^{s_i}$ in $I_{t+1}^{s_i}$, formally:
\begin{equation}
    P(r_{t+1}^{s_i} | I_1^{s_i}, I_2^{s_i}, I_3^{s_i}, ..., I_t^{s_i}, \{ I_{t+1}^{s_i}- r_{t+1}^{s_i} \}),
\end{equation}
Recently, many proposed KT methods studied the contextual information (i.e. side information) of the questions in KT problem~\cite{wang2019deep, pandey2020rkt, liu2019ekt, ghosh2020context}, and found that it can drastically help the KT model converge, suggesting that contextual information is strongly correlated to $r_{t+1}^{s_i}$. Consequently, we follow this work and give a more general representation of $I_t^{s_i}$ in KT problem, namely $I_t^{s_i} = (e_{t}^{s_i,q_j}, c^{q_j}, r_t^{s_i})$ where $c^{q_j}$ represents the contextual information of question $q_j$ (e.g. related concepts, part, etc.). Finally, $\kappa$ is defined as $\kappa_t^{s_i} = \{ s_i, q_j, c^{q_j}, r_t^{s_i} \}$, referring to all features that participated in one interaction $I_t^{s_i}$.

\subsection{Student-Related Features}
In order to achieve adaptive learning in DKT, the model needs to adapt itself to different students at different stages (i.e. student-related features). That is, when the model makes a personalized prediction, it must be aware of which student is being predicted and what characteristics this student currently has. Inspired by the observation that the proactive behavior sequence of each student actually reflects some of his inner properties, we argue that I. interactive sequences of different students at the same time period are distinguishable (distinctive), and II. interactive sequences of the same student at different time periods are also distinguishable, as long as it satisfies a. the features in an interaction is sufficient enough, b. the length of the interactive sequence is large enough, and c. the time interval of the same student's sequences is long enough. Accordingly, students' interactive sequences can be used to identify their own, and summarize the student-related features. As a result, a mild assumption was made for the latter analysis.
\begin{assumption}\label{assumpt:def}
For any interactive sequences satisfying $\left|\left| S_{t_0, t_1}^{s_i} \right|\right| > \Theta >> 1$, $ \left|\left| \kappa \right|\right| > \Psi$ and $t_2 - t_1 > \mathcal{E}$, $S_{t_0,t_1}^{s_i}$ can be distinguished from $S_{t_0,t_1}^{s_j}$ and $S_{t_2,t_3}^{s_i}$ respectively.
\end{assumption}
\section{Methodology}
\label{section:four}
In this section, we provided a comprehensive presentation of the proposed LANA method.
\subsection{Overview}
LANA method is composed of a LANA model and a training mechanism. The overall architecture of the proposed LANA model was presented in Figure~\ref{fig:lana-arch}. As it can be seen, the left part of the architecture in the figure is the encoder and the Student-Related Features Extractor (SRFE), while the right part is the decoder. The encoder aims to retrieve any useful information from the model's input embedding, then the SRFE further abstracts this information to obtain student-related features (Assumption~\ref{assumpt:def}). Finally, the decoder makes the prediction with the collected information from both SRFE and the encoder. LANA model, like SAINT+~\cite{shin2020saint+}, is a transformer~\cite{vaswani2017attention} based KT model. However, different from the SAINT+, the LANA model has mainly 3 improvements: I. LANA model considers KT's characteristics, and therefore makes modifications to the basic transformer model, such as feeding positional embedding directly into attention modules. II. LANA model uses a novel SRFE to abstract necessary student-related features from the input sequence. III. LANA model utilizes pivot module and extracted student-related features to dynamically construct different decoders for different students. With the reconstructed decoder, retrieved knowledge states, and other contextual information, corresponding personalized responses for future exercises are predicted.

\begin{figure*}[!t]
    \centering
    \includegraphics[width=0.85\linewidth]{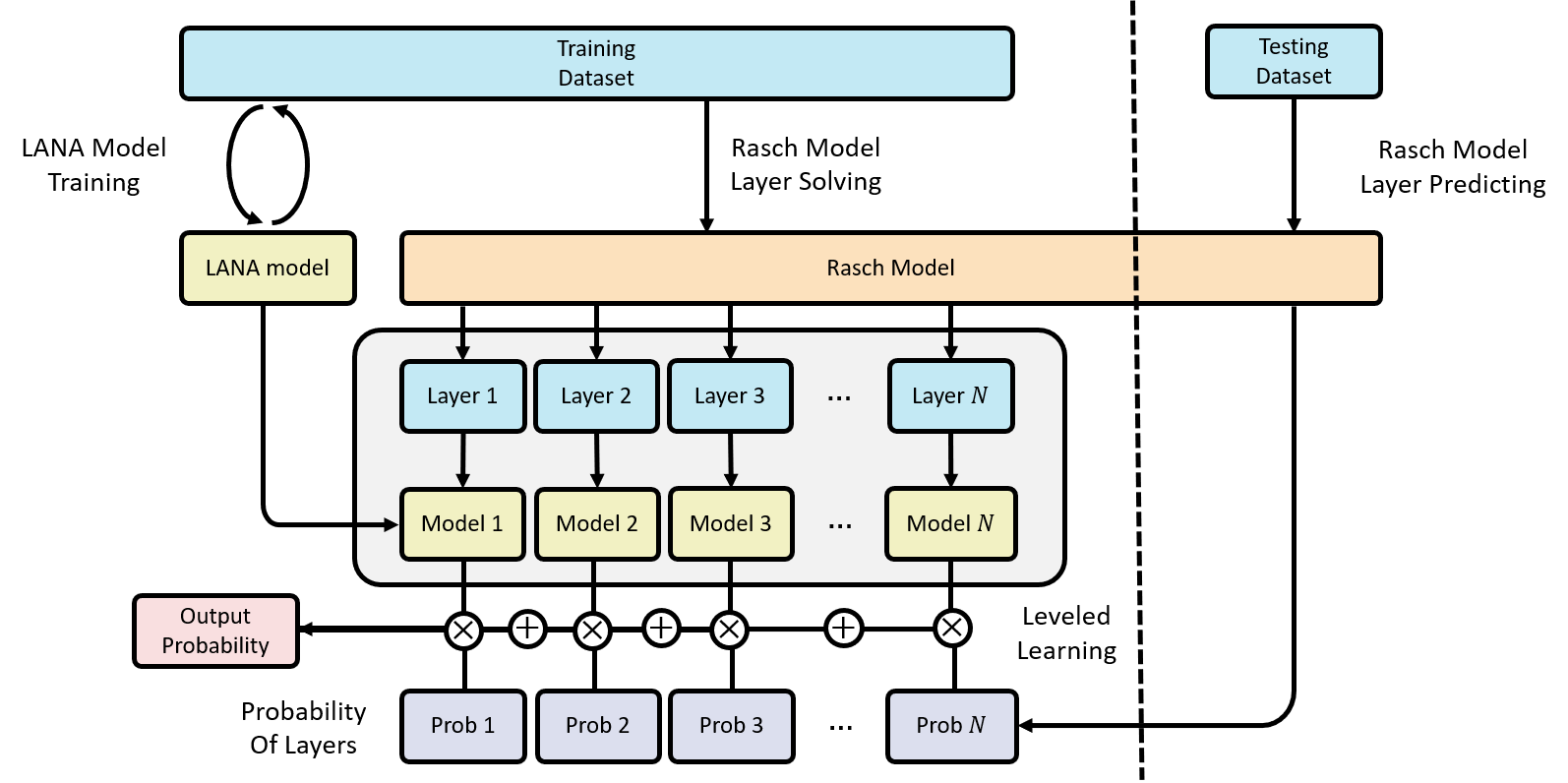}
    \caption{The workflow of leveled learning: interpretable Rasch model was leveraged to analyze students' overall ability levels, and then cluster students into multiple layers, where each layer would respectively fine-tune the LANA model by its own training data.}
    \label{fig:leveled-learning}
\end{figure*}

While the pivot module can help the LANA model transform the decoder in terms of students' inherent properties, the encoder, on the other hand, is fixed for all students after the training. Considering in DKT the input sequence is only a portion of the student's whole interactive sequence for lowering the computational requirement, the difficulty for features extracting from the input sequence increases, and therefore the ability for LANA model to distinguish different students at different stages needs to be strengthened greatly. Consequently, a leveled learning mechanism was proposed to address this problem by specializing different encoders and SRFEs to different groups of students with interpretable Rasch model defined student's ability level. The workflow of the leveled learning was illustrated in Figure~\ref{fig:leveled-learning}.
\subsection{Base Modifications}
\label{subsec:base-modification}
There are mainly two base modifications in the LANA model that were made to the basic transformer. Firstly, in the LANA model, the positional information (e.g. positional encoding, positional embedding) was directly fed into the attention module with a private linear projection, instead of being added to the input embedding and shared the same linear projection matrix with other features in the input layer. Although experiments in \cite{vaswani2017attention} suggested that blending input embedding with positional information is effective, recently some work~\cite{shiv2019novel} debated that when the model becomes deeper, it tends to ``forget'' the positional information fed into the first layer. Moreover, some other work~\cite{ke2020rethinking} believed that adding positional information to the input embedding and offering them to the attention module, is essentially making them share the same linear projection matrix, which is not reasonable since the effects of the input embedding and the positional information are clearly distinctive. For exactly the same reason, in the LANA model, multiple input embeddings (i.e. question ID embedding, student ID embedding, etc.) are concatenated instead of added, leading to the second base modification. Specifically, assumes there are $m$ input embeddings in total, each with a dimension of $D^f$. Then after concatenating, the input embedding would have a total dimension of $D^{mf}$. Hence, a $D^{mf} \rightarrow D^f$ linear projection layer was used to map the concatenated input embedding of dimension $D^{mf}$ to dimension $D^{f}$.

\begin{figure}[htb]
    \centering
    \includegraphics[width=0.8\linewidth]{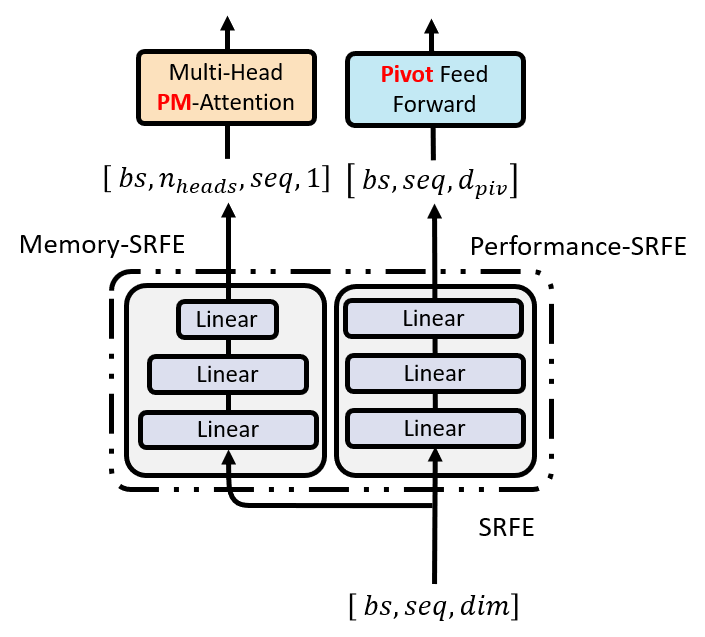}
    \caption{The data shape transformation of two SRFE: Memory-SRFE and Performance-SRFE.}
    \label{fig:srfe-arch}
\end{figure}

\begin{figure}[b]
    \centering
    \includegraphics[width=0.9\linewidth]{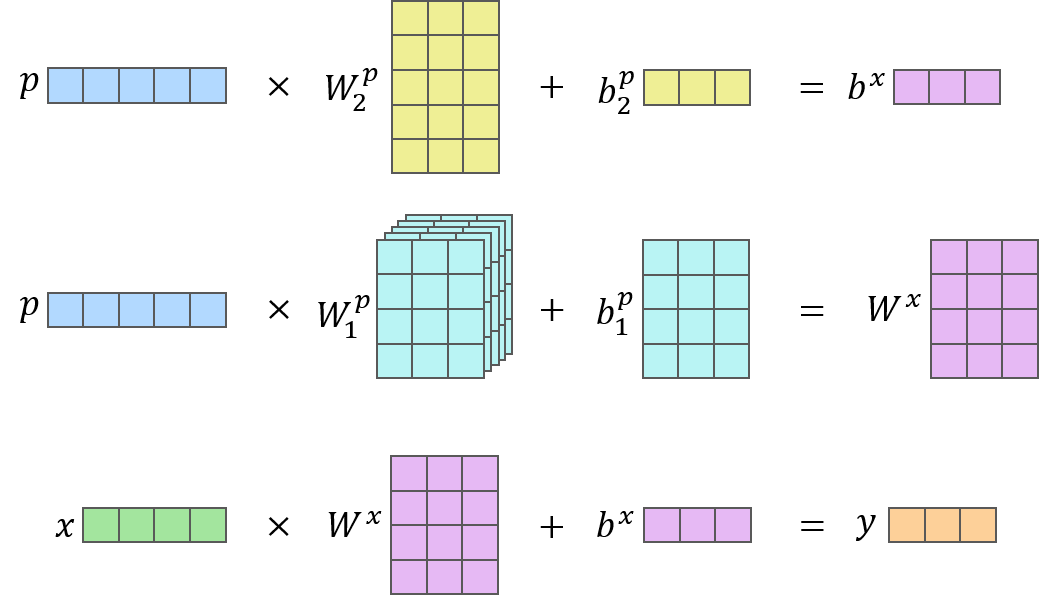}
    \caption{An illustration of the data transformation in the pivot module.}
    \label{fig:pivot-transform}
\end{figure}

\begin{figure}[t]
    \centering
    \includegraphics[width=0.4\linewidth]{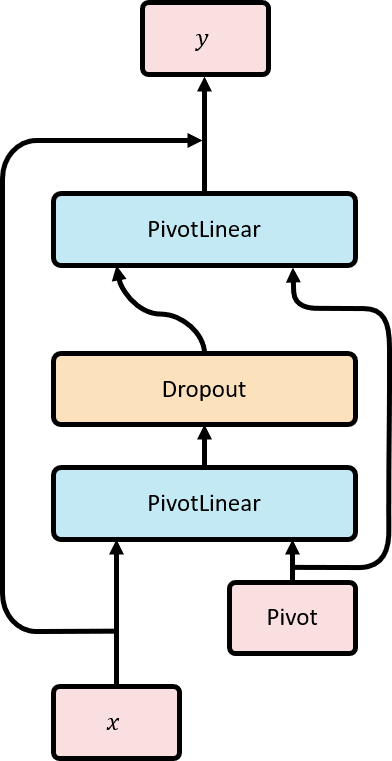}
    \caption{The inner structure of the Pivot Classification Feed Forward Network (PC-FFN) Module.}
    \label{fig:pc-ffn}
\end{figure}

\subsection{Student-Related Features Extractor (SRFE)}

Student-Related Features Extractor (SRFE) summaries students' inherent properties from their interactive sequences with Assumption~\ref{assumpt:def} for the pivot module to personalize the parameters of the decoder. Specifically, SRFE contains an attention layer and several linear layers, where the attention layer was used to distill student-related features from the provided information by the encoder, and the linear layers were leveraged to refine and reshape these features. It is notable that in the LANA model there were primarily two SRFEs: memory-SRFE and performance-SRFE, where the former one was utilized to derive students' memory-related features for the PMA module (be introduced later) and the latter one was dedicated to distill students' performance-related features (i.e. Logical thinking skill, Reasoning skill, Integration skill, etc.) for PC-FFN module (introduced later either). The reshaping process was drawn in Figure~\ref{fig:srfe-arch} for better illustration, where $bs$, $n_{heads}$, $seq$ and $d_{piv}$ are referred to as the model's batch size, the number of attention heads~\cite{vaswani2017attention}, the length of the input sequence and the dimension of performance-related features. The intuition that memory-related features have a second dimension of $n_{head}$ comes from the theory that each attention head only pays attention to one perspective of the features. Thus it is reasonable that each student has different memory skills for different attention heads (e.g. for different concepts).

\subsection{Pivot Module}
\label{subsec:pivot-module}
Provided an ordinary input $x$, a student-related features $p$ and a target output $y$, pivot module learns the process of learning how to project $x$ to $y$ based on $p$, instead of simply learning to project $x$ to $y$ (i.e. Pivot module learns to learn) as Equation~\ref{eq:pivot-1} shown.
\begin{equation}\label{eq:pivot-1}
    y = (f(p))(x),
\end{equation}
where $f(\cdot)$ here is the function that pivot module learns to learn. That is, the projection matrix of $x$ is adapted to $p$ instead of being fixed. To accomplish this dynamic mapping, the weight and bias of $x$ need to be a projection from $p$. Assumes $p \in \mathbb{R}^{D_p}$, $x \in \mathbb{R}^{D_x}$ and $y \in \mathbb{R}^{D_y}$, Equation~\ref{eq:pivot-1} could be formally presented in Equation~\ref{eq:pivot-3}:
\begin{equation}\label{eq:pivot-3}
    y = W^{x}x + b^{x},
\end{equation}
where $W^{x} \in \mathbb{R}^{D_y \times D_x}$ and $b^{x} \in \mathbb{R}^{D_y}$. Since $W^{x}$ and $b^{x}$ is derived from $p$, the detailed transformation could be revealed in Equation~\ref{eq:pivot-2}, which was also depicted in Figure~\ref{fig:pivot-transform} for better illustration.
\begin{equation}\label{eq:pivot-2}
    W^{x} = W_1^{p} p + b_1^{p},~~b^{x} = W_2^{p} p + b_2^{p},
\end{equation}
where $W_1^{p} \in \mathbb{R}^{(D_y \times D_x) \times D_p}$, $b_1^{p} \in \mathbb{R}^{(D_y \times D_x)}$, $W_2^{p} \in \mathbb{R}^{D_y \times D_p}$ and $b_2^{p} \in \mathbb{R}^{D_y}$.

By simplification, Equation~\ref{eq:pivot-1} can be defined as Equation~\ref{eq:pivot-4}, being named as $PivotLinear(x, p)$.
\begin{equation}\label{eq:pivot-4}
    y = (Wp)x + b = PivotLinear(x, p),
\end{equation}
where $W \in \mathbb{R}^{D_y \times D_x \times D_p}$ and $b \in \mathbb{R}^{D_y}$. 

In the LANA model, there are primarily two modules that pertain to the pivot module: Pivot Memory Attention (PMA) Module and Pivot Classification Feed Forward Network (PC-FFN) Module. In many methods~\cite{ghosh2020context, pandey2020rkt}, Vanilla Memory Attention (VMA) Module was employed to consider the ``forgetting'' behavior of students, which is pivotal in KT's context since students are very likely to have done similar exercises to the one he is going to do, and if the student could remember the answers to previous similar exercises, the probability of him correctly answering the future related exercises will be increased greatly. Inspired by the Ebbinghaus Forgetting Curve \cite{murre2015replication} and much previous work~\cite{pandey2020rkt, ghosh2020context}, ``forgetting'' behavior of students are defined as exponentially decaying weights of corresponding interactions in the timeline. Detailedly, in the original attention module, the weight of item $j$ on item $k$, i.e. $\alpha_{j, k}$, is determined by the \texttt{sigmoid} result of the similarity between item $j$ and item $k$:
\begin{equation}\label{eq:pivot-5}
    \alpha_{j, k} = \frac{sim(j, k)}{\sum_{k^{'}}sim(j, k^{'})},
\end{equation}
where $sim(\cdot)$ is a function to calculate the similarity between item $i$ and item $j$ by dot production. In order to take ``forgetting'' behavior into $\alpha_{j, k}$'s account (e.g. The further away from $j$, the lower the weight $\alpha_{j, k}$ would be), we replaced Equation~\ref{eq:pivot-5} with Equation~\ref{eq:pivot-6}:
\begin{equation}\label{eq:pivot-6}
    \alpha_{j, k, m} = \frac{e^{-(\theta + m) \cdot dis(j, k)} \cdot sim(j, k)}{\sum_{k^{'}}sim(j, k^{'})},
\end{equation}
where $m$ is the student's memory-related features extracted in memory-SRFE, $\theta$ is a private learnable constant that describes all students' average memory skill in the PMA module, and $dis(\cdot)$ calculates the time distance between item $j$ and item $k$ (e.g. item $j$ is done $dis(j, k)$ minutes after item $k$ is done). The reason for representing the memory skill with two learnable parameters is to reduce the difficulty for model converging since $m$ has a much longer back-propagation path compared to $\theta$. When $\theta$ is introduced to fit the average memory skill of all students, the distribution of $m$ becomes a Gaussian distribution, which makes the model much easier to learn.

On the other hand, PC-FFN was utilized to make the final prediction in reference to the performance-related features, and its architecture was illustrated in Figure~\ref{fig:pc-ffn}. The idea of this module comes from many investigations that the early layers in a deep neural network are often used as a feature extractor while the latter layers are often used as a decision-maker to decide which feature is useful to the output of the model. As a result, these investigations point out that many models are actually having similar early layers, and it is the latter layers that make these models distinctive in usage. Consequently, PC-FFN in the LANA model was utilized as a personalized decision-maker to adaptively make the final prediction based on students' distinctive inherent properties:
\begin{equation}
    PC-FFN(x,~p) = x + PivotLinear(PivotLinear(x,~p),~p),
\end{equation}
where $p$ is the students' performance-related features extracted in the performance-SRFE.

\subsection{Leveled Learning}
\label{subsec:leveled-learning}

While the pivot module enables the decoder to be transformable for different students, the encoder and the SRFE of the LANA model that provides necessary information for the pivot module remains the same for all students. This is not problematic if the length of the input sequence is large enough since Assumption~\ref{assumpt:def} assures long sequences are always distinguishable, unless they both belong to the same student at the same time period. However, DKT, especially transformer-based DKT, can only be inputted with the latest $n$ (commonly $n = 100$) interactions at once due to the limited memory size and high computational complexity. Consequently, it is possible for the encoder and SRFE to output similar results for two different students, resulting in a failure for the decoder to adapt. To alleviate this problem, it is natural to think of assigning different students with different encoders and SRFEs that are highly specialized (sensitive) to their assigned students' patterns. However, in practice, it is not feasible to train a unique encoder for each individual student considering both the limited training time and the limited training data. As a result, a novel leveled learning method was proposed to address this problem, which was initially inspired by the fine-tuning mechanism in transfer learning~\cite{tan2018survey}, where we consider each student a unique task, and we want to transfer a model that fits well on all students to one student $s_i$ efficiently.

Leveled learning holds the view that the earlier layers of a model are similar for similar tasks. Thus, to save training time and enlarge the training set, instead of training each student a unique encoder and SRFE by his private training data, students with similar ability levels are considered to be grouped together, sharing their private training data and having the same encoder and SRFE. Therefore, LANA firstly utilizes an interpretable Rasch model to analyze the ability level $a^{s_i}$ for each student $s_i$, then groups students into different independent layers $l_i$. Assuming the ability distribution of all students and students at the level $l_i$ are Gaussian distribution $N(\mu_a, \sigma_a^2)$ and $N(\mu_i, \sigma_i^2)$ respectively, we have the Equation~\ref{eq:ll-1}:
\begin{equation}\label{eq:ll-1}
    \mu_a = \frac{\sum_i \mu_i}{L},~~\sigma_a^2 = \sum_i \sigma_i^2.
\end{equation}
In LANA, for simplicity, we consider all layers share the same variance $\sigma^2$ \footnote{In practice, if the number of layers is small, their variances then need to be manually measured and tuned based on the targets. If the number of layers is large, then multiple layers can be regarded as one layer and therefore sharing the same variance for all layers should be fine.}, and the difference of mean $\mu_i$ between consecutive layers is a constant $\tau$. Hence, $\mu_i$ and $\sigma_i^2$ are given by:
\begin{equation}\label{eq:ll-2}
    \mu_i = \mu_a - \frac{L - 1}{2} \times \tau + i \times \tau,~~\sigma_i^2 = \frac{\sigma_a^2}{L}.
\end{equation}
where $L = \left|\left|l_i\right|\right|$ is the number of layers. With both $\mu_i$ and $\sigma_i^2$ retrieved for every layer $l_i$, given a student's ability constant $a^{s_i}$, we can now calculate the probability of $s_i$ been grouped into different layers by Equation~\ref{eq:ll-3}:
\begin{equation}\label{eq:ll-3}
    p_i^{s_i} = \frac{\phi_{i}(a^{s_i})}{\sum_{i^{'}} \phi_{i^{'}}(a^{s_{i^{'}}})},~~\phi_{i}({a^{s_i}}) = \frac{1}{\sigma_i \sqrt{2 \pi}} e^{-\frac{(a^{s_{i}} - \mu_i)^2}{2 \sigma_i^2}}
\end{equation}
where $p_i^{s_i}$ is referred as the probability of student $s_i$ being grouped into layer $l_i$. As it can be seen from Equation~\ref{eq:ll-3}, students that have high ability levels are not necessarily grouped into layers with high expected ability levels $\mu_i$. Contrarily, these high ability students only have a higher probability of been grouped into high ability layers in comparison with those low ability students, which obeys rules in reality (e.g. high ability students may also come from normal schools).

Then, the LANA model that has been pre-trained on all students was duplicated $L$ times, each cloned model $m_i$ would be assigned to a layer $l_i$ to be dedicatedly fine-tuned with $l_i$'s private training data by weighted back-propagation:
\begin{equation}\label{eq:ll-4}
    loss_i = p_i * loss(predict_i, target),
\end{equation}
where $predict_i$ is the prediction of the model $m_i$.

While the training phase of leveled learning seems promising, the inference phase of it suffers problems. The first problem is how to make the prediction using multiple specialized models. In LANA, the prediction was made by $top-k$ models fusion. Detailedly, when student $s_i$'s future responses are needed to be predicted, LANA firstly computes $p_i$, then feed $s_i$'s interactive sequence to all models $m_i$ that satisfies $p_i \in top-k(p)$, where $k$ needs to be manually set up to control the predicting time. Then, the outputs of these models would be multiplied by $sigmoid(p_i)$ to form the final prediction. The workflow of leveled learning's inference step could be described in Equation~\ref{eq:ll-5}:
\begin{equation}\label{eq:ll-5}
    r_i = \sum_{i^{'}} (m_i(x) \times \sum_{h \in i^{'}} \frac{p_h}{\sum_{h^{'}} p_{h^{'}}}),~~i^{'} \in \{~i~|~p_i \in top-k(p)~\},
\end{equation}
where $r_i$ is the leveled learning's final prediction and $x$ is the input of the model. This workflow seems similar to the ensemble where multiple models are unitized to generate the final answer. Nonetheless, weights of models in LANA are probabilities that come from an interpretable Rasch model so that it is clear which model is dominant to $x$. Moreover, unlike in ensemble, where the role of each model is ambiguous, in LANA, every model has its explainable effect (e.g. $l_L$ is committed to high ability students, and therefore a student with large $p_L$ indicates he must be similar to those high ability students in $l_L$), suggesting that leveled learning significantly outperforms ensemble in interpretability. Detailed comparison was shown in Table~\ref{tab:compare-ll-ensemble}.

\begin{table}[htbp]
  \centering
  \captionsetup{justification=centering}
  \caption{\\\textsc{Comparison Between Leveled Learning And Ensemble}}
    \begin{tabular}{lll}
    \toprule
    \toprule
          & Leveled Learning & Ensemble \\
    \midrule
    Sub-set Select & Psycho-statistics & Random \\
    Interpretability & Good  & Bad \\
    Predicting Time & Controllable (top-k) & Uncontrollable \\
    \bottomrule
    \bottomrule
    \end{tabular}%
  \label{tab:compare-ll-ensemble}%
\end{table}%

On the other hand, the second problem of the leveled learning is how to compute $p_i$ for students that LANA has never met in training, namely the ``cold start'' problem~\cite{wilson2016estimating}. In vanilla KT context, we can only initiate newly arrived students' ability levels to the average ability level of all students. However, in practice, we can estimate their ability levels more accurately by asking them to do a couple of sample exercises or using ranking at school.

\section{Experiments}
\label{section:five}

\subsection{Experimental Setup}

In order to evaluate the effectiveness of the proposed LANA, we applied it to two real-world large-scale datasets in comparison with many other State-Of-The-Art (SOTA) KT methods. Specifically, EdNet~\cite{choi2020ednet} and RAIEd2020~\cite{riiid2020rac} are employed in our experiments, where EdNet is currently the largest publicly available benchmark dataset in education domain, consisting of over 90,000,000 interactions and nearly 800,000 students. On the other hand, RAIEd2020 is a recently published real-world dataset that has approximately the same size as EdNet with nearly 100,000,000 interactions and 400,000 students. Particularly, the average number of exercising interactions per student in RAIEd2020 is double to EdNet's. Other details of these two datasets can be referred to in Table~\ref{tab:dataset-detail}.
\begin{table}[t]
  \centering
  \captionsetup{justification=centering}
  \caption{\\\textsc{Datasets Comparison Of EdNet~\cite{choi2020ednet} And RAIEd2020}}
    \begin{tabular}{ccc}
    \toprule
    \toprule
    Statistic & EdNet & RAIEd2020 \\
    \midrule
    \# of interactions & 95,293,926 & 99,271,300 \\
    \# of students & 784,309 & 393,656 \\
    \# of exercises & 13,169 & 13,523 \\
    \# of parts & 7     & 7 \\
    \bottomrule
    \bottomrule
    \end{tabular}%
  \label{tab:dataset-detail}%
\end{table}%
\begin{table}[htbp]
  \centering
  \captionsetup{justification=centering}
  \caption{\\\textsc{The Detailed Hyper-parameters Setup Of Experiments Under Two Different Datasets}}
    \begin{tabular}{ccc}
    \toprule
    \toprule
          & EdNet & RAIEd2020 \\
    \toprule
    Optimizer & AdamW & AdamW \\
    Learning Rate & 5.00E-04 & 5.00E-04 \\
    Batch Size & 256   & 256 \\
    Input Length & 100   & 100 \\
    Model Dimension & 512   & 256 \\
    Hidden Dimension & 512   & 256 \\
    \# of Heads & 8     & 8 \\
    \# of Encoder & 4     & 2 \\
    \# of Decoder & 4     & 2 \\
    Dropout & 0   & 0 \\
    $\tau$ & 1.0 & 1.0 \\
    \bottomrule
    \bottomrule
    \end{tabular}%
  \label{tab:hyper-setup}%
\end{table}%
Moreover, 6 KT methods that had previously achieved SOTA performance have participated in the comparison, which would be respectively introduced in the following literature:
\begin{itemize}
    \item DKT~\cite{piech2015deep}: DKT models students' knowledge states and makes future response predictions with LSTM~\cite{hochreiter1997long}. Expressly, the knowledge states of students in DKT are represented by the hidden vectors of the LSTM.
    \item DKVMN~\cite{zhang2017dynamic}: DKVMN utilizes two novel matrices, a key matrix and a value matrix, to separately store each question's knowledge concepts and each student's mastery level of the corresponding concept.
    \item SAKT~\cite{pandey2019self}: SAKT replaces DKT's LSTM module with transformer's~\cite{vaswani2017attention} self-attention module to achieve SOTA on multiple knowledge tracing datasets.
    \item SAINT~\cite{choi2020towards}: While SAKT only uses one self-attention module in its architecture, SAINT leverages the whole transformer, which drastically improves the prediction accuracy.
    \item SAINT+~\cite{shin2020saint+}: SAINT+ is a follow-up of SAINT, where SAINT+ adds two more features to the input of the model. Experiments demonstrate the effectiveness of SAINT+ compared to SAINT.
    \item AKT~\cite{ghosh2020context}: AKT utilizes a monotonic attention mechanism to trace students' knowledge states, which considers the ``forgetting'' behavior of the human brain.
\end{itemize}
\begin{table}[t]
  \centering
  \captionsetup{justification=centering}
  \caption{\\\textsc{The AUC Comparison Of Different Methods Tested On EdNet And RAIEd2020 datasets}}
    \begin{tabular}{ccc}
    \toprule
    \toprule
    Dataset & Model & AUC \\
    \midrule
    EdNet & DKT   & 0.7638$^r$ \\
    EdNet & DKVMN & 0.7668$^r$ \\
    EdNet & SAKT  & 0.7663$^r$ \\
    EdNet & SAINT & 0.7816 \\
    EdNet & SAINT+ & \textit{0.7913} \\
    EdNet & SAINT+ \& \textit{BM} & 0.7935 \\
    EdNet & LANA  & \textbf{0.8059} \\
    \midrule
    RAIEd2020 & SAKT  & 0.7832 \\
    RAIEd2020 & AKT   & 0.7901 \\
    RAIEd2020 & SAINT+ & \textit{0.7956} \\
    RAIEd2020 & SAINT+ \& \textit{BM} & 0.7991 \\
    RAIEd2020 & LANA  & \textbf{0.8056} \\
    \bottomrule
    \bottomrule
    \end{tabular}%
  \label{tab:auc-comparison}%
\end{table}%
\begin{table*}[htb]
  \small
  \centering
  \caption{\\\textsc{Investigation Of The Effectiveness Of Different Improvements In LANA}}
    \begin{tabular}{cccccccc}
    \toprule
    \toprule
    \multirow{2}[2]{*}{Dataset} & \multirow{2}[2]{*}{Base Modification} & \multicolumn{2}{c}{Pivot Module} & \multirow{2}[2]{*}{Layered Learning} & \multirow{2}[2]{*}{AUC} & \multirow{2}[2]{*}{Improvement} \\
    \cmidrule{3-4} & & PMA & PC-FFN &  &  \\
    \midrule
    EdNet &       &        &       &       & 0.7913  & - \\
    EdNet & $\checkmark$     &        &       &       & 0.7935 & $\uparrow$~0.0022  \\
    EdNet &      & $\checkmark$        &       &       & 0.7997 & $\uparrow$~0.0084  \\
    EdNet &     &      & $\checkmark$    &     & 0.7923  & $\uparrow$~0.0010 \\
    EdNet &     &      &    & $\checkmark$     & 0.7933  & $\uparrow$~0.0020 \\
    EdNet & $\checkmark$     & $\checkmark$        &       &       & 0.8029  & $\uparrow$~0.0116  \\
    EdNet &      & $\checkmark$     & $\checkmark$ &       & 0.8015  & $\uparrow$~0.0102  \\
    EdNet & $\checkmark$     & $\checkmark$     & $\checkmark$ &       & 0.8038  & $\uparrow$~0.0125  \\
    EdNet & $\checkmark$     & $\checkmark$     &   & $\checkmark$     & 0.8050  & $\uparrow$~0.0137 \\
    EdNet & $\checkmark$     & $\checkmark$     & $\checkmark$   & $\checkmark$     & \textbf{0.8059}  & $\uparrow$~\textbf{0.0146} \\
    \midrule
    RAIEd2020 &       &       &       &       & 0.7956 & - \\
    RAIEd2020 & $\checkmark$     &       &       &       & 0.7991 & $\uparrow$~0.0035 \\
    RAIEd2020 &      & $\checkmark$       &       &       & 0.8020 & $\uparrow$~0.0064 \\
    RAIEd2020 &        &       & $\checkmark$       &      & 0.7965 & $\uparrow$~0.0009 \\
    RAIEd2020 &        &       &       & $\checkmark$     & 0.7977 & $\uparrow$~0.0021 \\
    RAIEd2020 & $\checkmark$     & $\checkmark$       &       &       & 0.8031 & $\uparrow$~0.0075 \\
    RAIEd2020 &       & $\checkmark$      & $\checkmark$ &       & 0.8027  & $\uparrow$~0.0071 \\
    RAIEd2020 & $\checkmark$     & $\checkmark$     & $\checkmark$ &        & 0.8035  & $\uparrow$~0.0079 \\
    RAIEd2020 & $\checkmark$     & $\checkmark$     &   & $\checkmark$     & 0.8051   & $\uparrow$~0.0095 \\
    RAIEd2020 & $\checkmark$        & $\checkmark$       & $\checkmark$       & $\checkmark$     & \textbf{0.8056}  & $\uparrow$~\textbf{0.0100} \\
    \bottomrule
    \bottomrule
    \end{tabular}%
  \label{tab:ablation-study}%
\end{table*}%
\begin{figure}[htb]
    \centering
    \subfloat[]{%
        \includegraphics[width=0.46\linewidth]{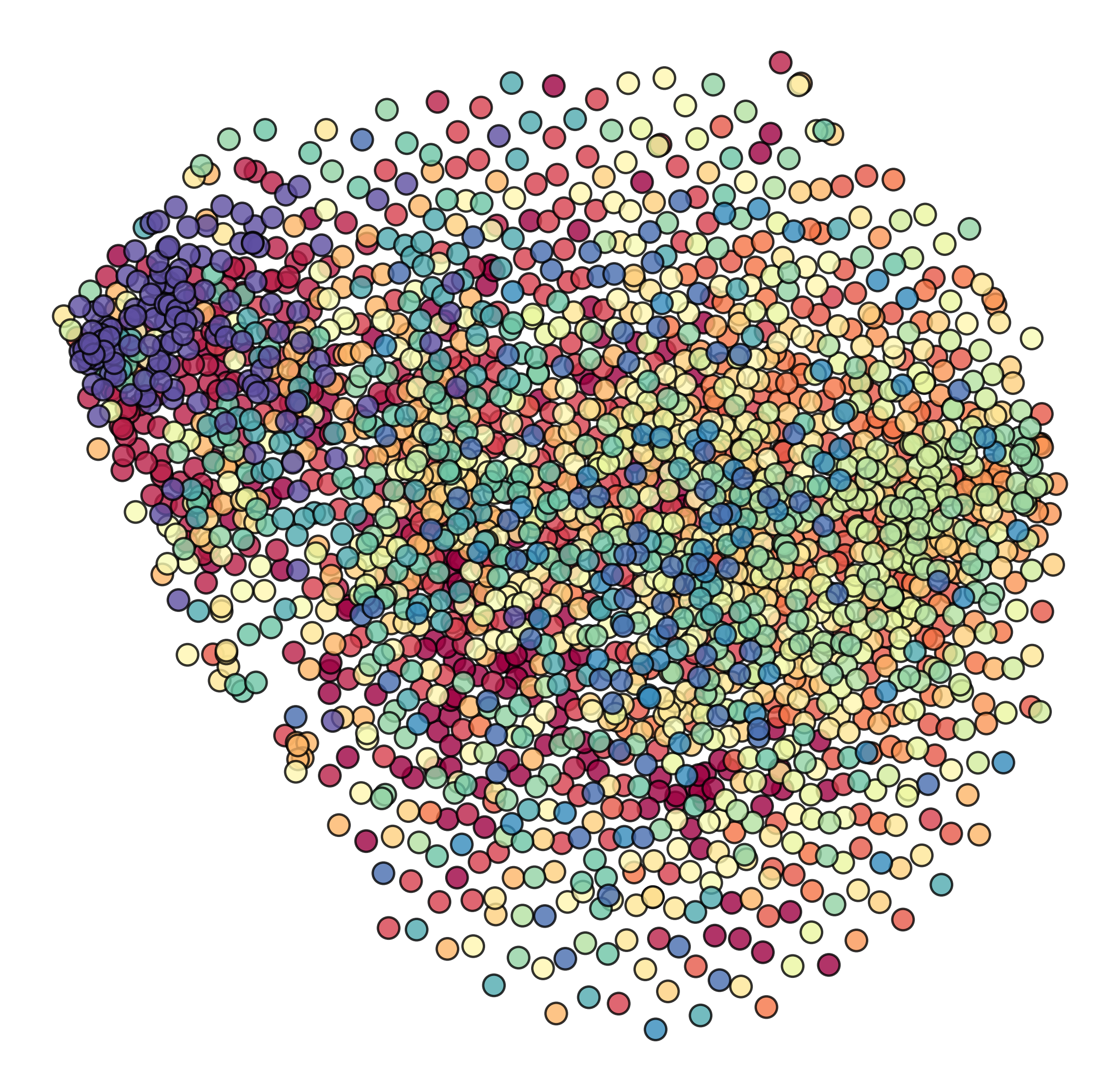}}
    \hfill
    \subfloat[]{%
        \includegraphics[width=0.54\linewidth]{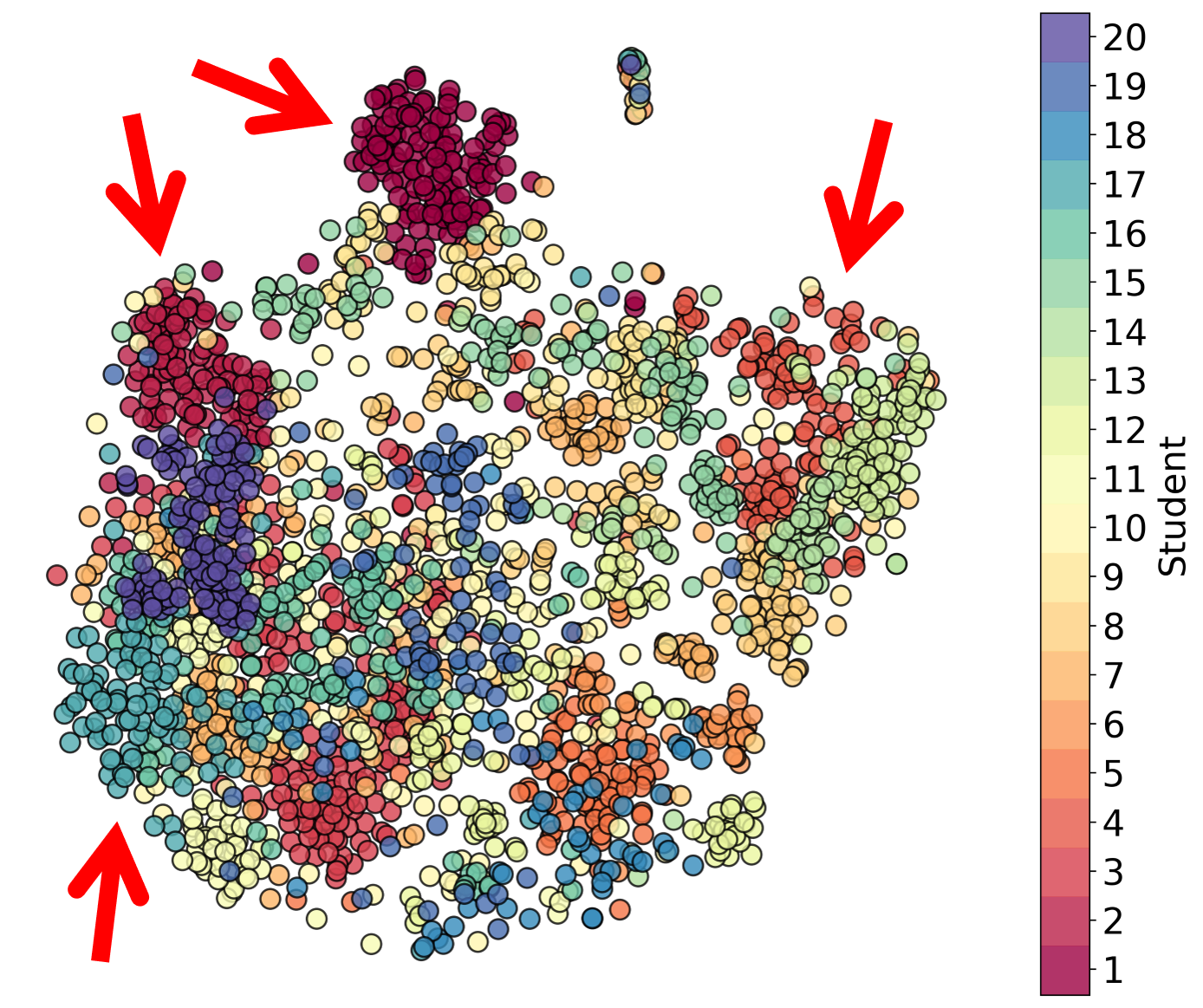}}
    \\
    \subfloat[]{%
        \includegraphics[width=0.45\linewidth]{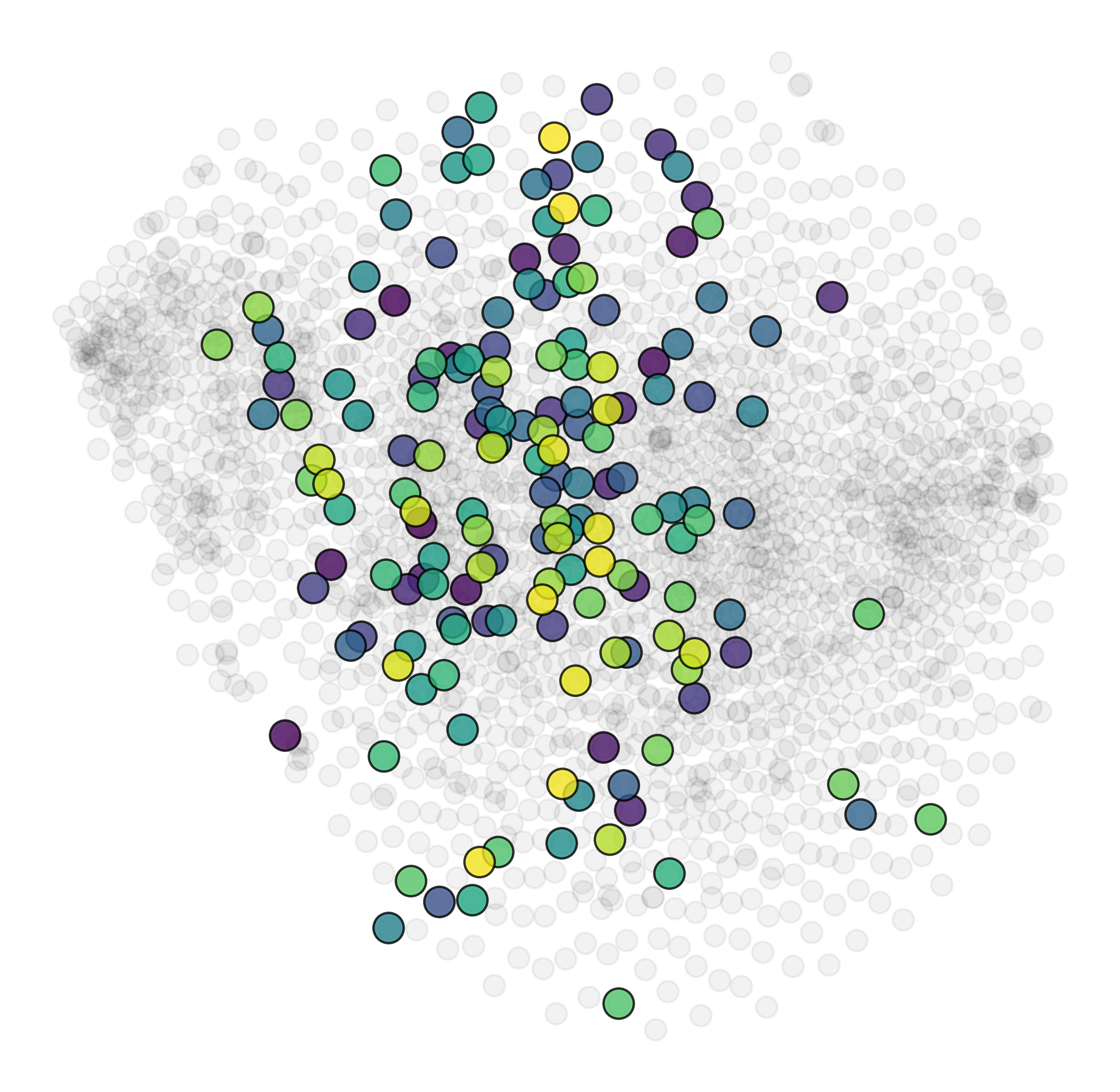}}
    \hfill
    \subfloat[]{%
        \includegraphics[width=0.55\linewidth]{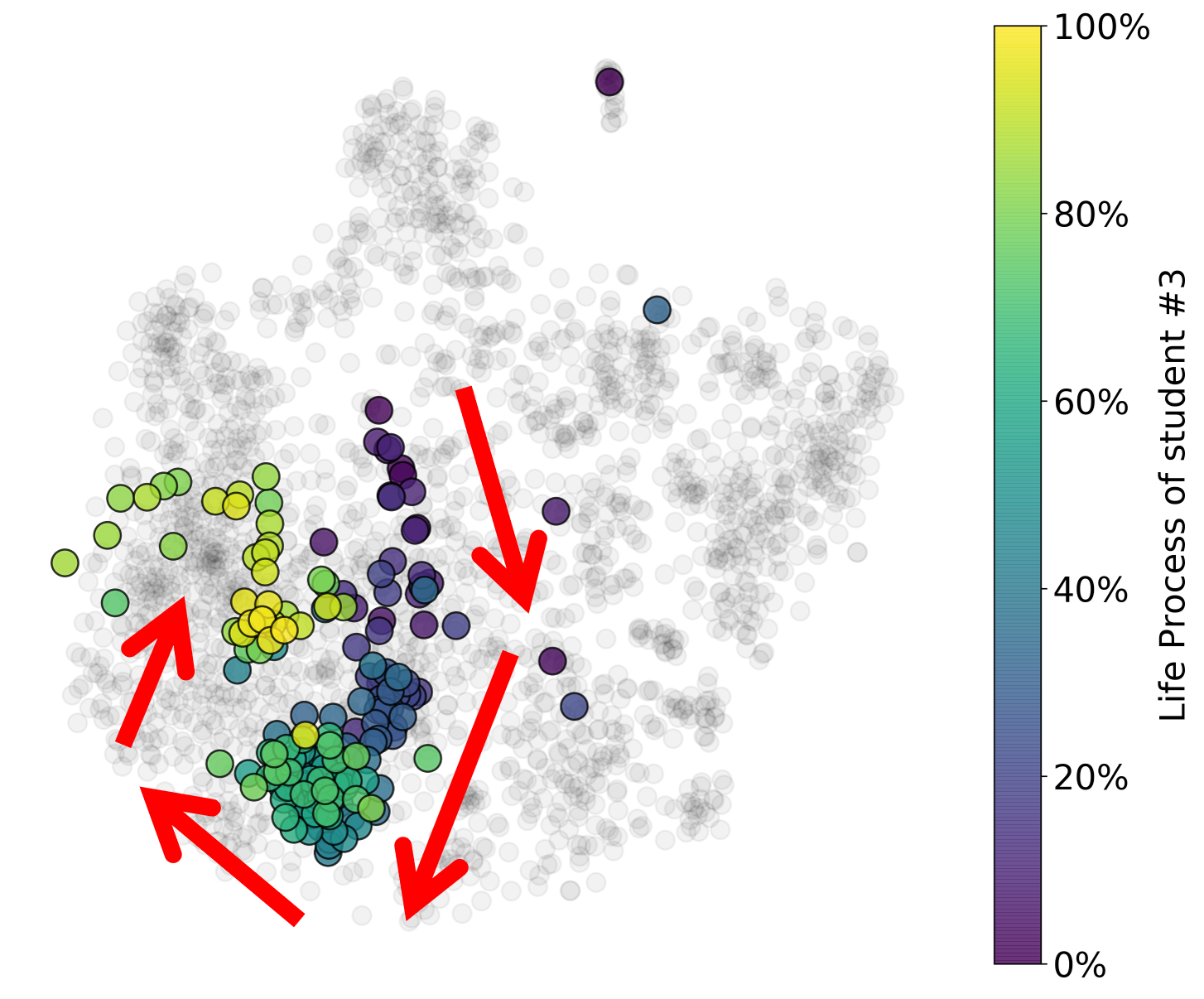}}
    \caption{The visualization of intermediate features in SAINT+ (a) and in LANA (b). Compared to (a), students in (b) (different colors) are notably clustered (marked arrows). The learning process of student \#3 overtime in SAINT+ (c) and in LANA (d). compared to (c), a clear learning path appeared in (d).}
    \label{fig:decoder-visualization}
\end{figure}

In terms of the basic experimental environment, all experiments were conducted with Pytorch\footnote{\url{https://pytorch.org/}} 1.6 on a Linux server that is equipped with an Nvidia V100 GPU. For hyper-parameters setup, the learning rate was set to $5e-4$ with $AdamW$~\cite{loshchilov2017decoupled} optimizer, the length of the input sequence was set to $100$, the batch size was set to $256$, and other detailed configurations were listed in Table~\ref{tab:hyper-setup}. The input features $\kappa$ in EdNet contains \textit{Question ID}, \textit{Question part}, \textit{Students' responses}, \textit{Time interval between two consecutive interactions} and \textit{Elapsed time of an interaction}, whereas in RAIEd2020, a new feature is additionally added to $\kappa$, which indicates \textit{Whether or not the student check the correct answer to the previous question}. Finally, The Area Under the receiver operating characteristic Curve (AUC) was leveraged in our experiments as the performance metric, which has been widely used in many other KT-related proposals.

For the ease of explanation, hereinafter Base Modification (Section~\ref{subsec:base-modification}), Pivot Module (Section~\ref{subsec:pivot-module}) and Leveled Learning (Section~\ref{subsec:leveled-learning}) would be abbreviated as \textit{BM}, \textit{PM} and \textit{LL} respectively.

\subsection{Results And Analysis}

The overall experimental results of different KT methods on different datasets were illustrated in Table~\ref{tab:auc-comparison}. Because we had successfully reproduced the performance of SAINT and SAINT+ that was previously reported in SAINT+'s paper~\cite{shin2020saint+} (with considerable precision), AUCs of other models are therefore directly cited from the paper (labeled with subscript $r$).

From the comparison table, it can be seen that in both EdNet and RAIEd2020 datasets, LANA (marked bold) outperforms the previous SOTA method (marked italic) by 1.46\% and 1.00\% respectively, readily verifying the effectiveness of our proposed improvements. Moreover, LANA also surpasses SAINT+ \& \textit{BM} by 1.24\% and 0.65\% respectively, suggesting adaptability contributes most to LANA's AUC increment. Considering experimented datasets are by far the two largest knowledge tracing datasets in the world, these results undoubtedly provide strong evidence of the validity of the proposed LANA method. 

\subsection{Ablation Studies}

In this section, we investigated the effectiveness of each of our proposed improvements: \textit{BM} that customizes the basic transformer architecture, \textit{PM} that enables the decoder to be adaptive to the students' personal characteristics, and \textit{LL} that interpretably specializes encoders and SRFEs for better predicting performance. The results of the ablation study were shown in Table~\ref{tab:ablation-study}.

The table shows in EdNet, applying \textit{BM} alone was already capable of improving the predicting AUC by approximately 0.2\% averagely, verifying the importance of both the action of positional embedding and the personalized linear projection for each input feature in KT's context. Meanwhile, applying \textit{LL} solely can benefit the model performance as well, by generally 0.2\% compared to 0.1\% with the vanilla ensemble. Considering without \textit{PM}, \textit{LL} would just perform fitting on students with different ability levels, the performance gain from sole \textit{LL} could be interpreted as reductions in students' inherent properties gaps. Moreover, \textit{BM + PM} drastically boosts the model performance by nearly 1.25\%, suggesting \textit{PM} makes proper use of extracted student-related features from SRFE to adaptively reparameterize the model's decoder for different students at different stages, and therefore contributes most to the final performance gain. Finally, by combining all improvements together, \textit{BM + PM + LL} (i.e. LANA) achieves a final AUC of 0.8059, substantially outperforms previous SOTA by at least 1.46\%.

\subsection{Features Visualization}

For vividly illustrating the validity of student-related features distilling in LANA, $20$ students' intermediate features from PC-FFN module was sampled to generate Figure~\ref{fig:decoder-visualization} by t-SNE\cite{van2008visualizing}. In figure~\ref{fig:decoder-visualization} (a) and (b), each sample represents intermediate features of different students with different colors in SAINT+ and LANA respectively. It can be seen that in SAINT+, samples are almost randomly distributed, indicating the correlation between samples of the same student is not more significant compared to samples of the others due to the ignorance of students' personalities. On the other hand, in LANA, clusters (marked arrows) of samples have notably appeared in comparison to (a). Thus, we concluded that LANA is capable of successfully extracting student-related features from their interactive sequences, summarizing the similarities and differences, which eventually results in more distinguishable features for the final classifier.

Furthermore, we individually visualized student \#3's (randomly picked) samples along the time axis to investigate the transitioning pattern of features in Figure~\ref{fig:decoder-visualization} (c)(SAINT) and (d)(LANA). In (c), there is no clear pattern in the change of features over time, while in (d), a clear transitioning path could be noticed. Since many other students are sharing the same pattern in LANA, we argue that it represents the trajectory of the student's ability changes with more and more exercising. Namely, it is the learning path of the student. Consequently, we contended that it is potentially helpful for other applications, such as learning stages transfer and learning path recommendation.

\section{Conclusion}
\label{section:six}

Knowledge Tracing, as a fundamental task to achieve adaptive learning, requires the tracing method capable of providing personalized analysis for each individual student. However, while integrating deep learning with knowledge tracing creates the most powerful DKT that surpasses any other traditional KT methods in metrics, the adaptability tends to be ignored by DKT and its variants considering they are essentially treating all students averagely to learn the pattern. 

In this paper, we proposed a novel \textbf{L}eveled \textbf{A}ttentive K\textbf{N}owledge Tr\textbf{A}cing (LANA) method that was committed to bringing adaptability back to DKT. Specifically, inspired by BKT and IRT, LANA aims to achieve adaptability by providing different model parameters for different students at different stages. However, separately training massive unique models for each individual student is clearly not a practical solution, and therefore instead of directly learning the model parameters of different students, LANA distills students' inherent properties from their respective interactive sequences by a novel SRFE, and learns the function to reparameterize the model with these extracted student-related features. Consequently, innovative pivot module was proposed to produce an adaptive decoder. Besides, to reduce the ambiguity of the input sequence and capture the long-term characteristic of the individual student, a novel leveled learning training mechanism was introduced to cluster students by interpretable Rasch model defined ability level, which not only specializes the encoder and therefore enhances the significance of students' latent features, but also saves much training time. Extensive experiments on the two largest public benchmark datasets in the education domain strongly evaluate the feasibility and effectiveness of the proposed LANA. Features visualization also suggests extra impacts of LANA, be it learning stages transfer or learning path recommendation. 

However, there are some drawbacks of LANA as well: firstly, the configuration of variances in Layered Learning requires extensive manpower, which is desired to have a automatic workflow to setup these parameters. Secondly, while Figure~\ref{fig:decoder-visualization} illustrates the efficacy of the proposed SRFE module, a more systematic way to quantify the student features needs to be came up with. Consequently, solving these problems will be our next step. 




%
\bibliographystyle{abbrv}

\balancecolumns
\end{document}